\documentclass[aps,prl,reprint,superscriptaddress]{revtex4-2}
\usepackage{graphicx}
\usepackage{dcolumn}
\usepackage{bm}
\usepackage{amsmath}
\usepackage{amssymb}
\usepackage{braket}

\begin{document}
\title{Microwave Dressed States and Vacuum Fluctuations in a  Superconducting Condensate}
\author{Anoop Dhillon}
    \email{Contact author: a33dhillon@uwaterloo.ca}
    \affiliation{Department of Electrical and Computer Engineering, University of Waterloo, Waterloo, Ontario, Canada}
\author{A. Hamed Majedi}
    \email{Contact author: ahmajedi@uwaterloo.ca}
    \affiliation{Department of Electrical and Computer Engineering, University of Waterloo, Waterloo, Ontario, Canada}
    \affiliation{Department of Physics and Astronomy, University of Waterloo, Waterloo, Ontario, Canada}
    \affiliation{Waterloo Institute of Nanotechnology, Waterloo, Ontario, Canada}
\date{\today}

\begin{abstract}
% Abstract
Microwave dressed states are found to emerge within the superconducting condensate when coupled to a quantized electromagnetic field due to photon-Cooper pair entanglement. The renormalized energy separation between these states exceeds the prediction of BCS theory, with the enhancement depending on the number of photons and also arising from electromagnetic vacuum fluctuations. Our work introduces an equilibrium quantum model of microwave-enhanced superconductivity, expanding the theoretical description beyond Eliashberg’s non-equilibrium theory. We further demonstrate that the superconducting condensate exerts a back-action on the electromagnetic field, suppressing electric field fluctuations, including those from the vacuum state. This result is consistent with Glauber and Lewenstein's field quantization in dielectric media.

\end{abstract}

\maketitle

% Introduction
The interaction between superconductors and electromagnetic (EM) fields has long been of interest to the fields of microwave electronics \cite{multiphotonProcTunneling, quantumDetectionMMWavelength, weinstock2012microwave, BraginskiSuperconductorElectronicsStatus2019, Majedi-IFE}, non-equilibrium superconductivity \cite{W.D.M+MicrowaveEnhancedCriticalSupercurrents1966, L.NMechanismSuperconductivityStimulated1976, D.KNonequilibriumSuperconductivityMicrowave1979}, and superconducting quantum computing \cite{GarciaRipollQuantumInformationQuantum2022, PhysRevA.69.062320}. While these fields rely on an understanding of how classical and quantized electromagnetic fields couple to superconductors, a comprehensive quantum optical theory of this interaction remains unexplored.

In this paper, we investigate the direct quantization of electromagnetic fields within a superconducting condensate, treating the two as a coupled, bipartite quantum system. We show that microwave-dressed states emerge via photon-Cooper pair entanglement, and that the renormalized energy of the system depends not only on superconducting parameters, such as the energy gap, but also on the number of photons present. While related techniques are common in circuit quantum electrodynamics (QED) \cite{PhysRevA.69.062320, GarciaRipollQuantumInformationQuantum2022}, to our knowledge, they have not been applied to the superconducting condensate in isolation.

We find that the energy difference between these dressed states exceeds that predicted by BCS theory \cite{PhysRev.108.1175}, and varies with the strength of the applied field. This mechanism enables microwave-enhanced superconductivity at equilibrium, offering a significant departure from Eliashberg theory \cite{Eliashberg1971,osti_4687835}, which relies on non-equilibrium superconducting dynamics to suppress pair-breaking processes through the redistribution of quasiparticles under high-intensity fields. Notably, the effect persists under low-power conditions and is even induced by electromagnetic vacuum fluctuations alone.

Several recent works have explored related concepts, showing that the superconducting gap can be modified when low-dimensional materials are placed in external electromagnetic cavities. These studies attribute this gap enhancement to phonon-mediated effects \cite{cavityEliashbergEnhance, 10.1063/5.0231202, L.S.S+CavityenhancedSuperconductivityMgB22024}, modified electronic dispersion \cite{PhysRevB.111.035410}, or pair-density wave formation \cite{cavityMediatedSC}. While these works provide valuable insights, they are constrained by material-specific assumptions and cavity configurations. In contrast, we propose a generalized, material-independent mechanism that does not rely on an engineered photonic environment. Instead, we show that photon-condensate entanglement intrinsically alters the energy spectrum of any superconducting system, revealing a universal interaction with meaningful implications for superconducting device performance, noise, and coherence.

Beyond modifying the condensate, our quantum optical framework enables the examination of its back-action on the quantized electromagnetic field via the minimal coupling approximation. We show that electric field fluctuations inside the condensate are suppressed relative to free space, consistent with Glauber and Lewenstein's result for dielectric media \cite{G.LQuantumOpticsDielectric1991}. This reflects the condensate's effective behaviour as a dielectric with negative, frequency-dependent permittivity below its energy gap \cite{mei1991electromagnetics}. The remainder of this paper presents our theoretical framework and explores the implications of these results.

% Derivation
We begin by writing the Hamiltonian density of the condensate in the presence of the quantized electromagnetic field as,
\begin{align}
    \label{eqn:fullHamiltonian_nonquantized}
    \mathcal{H}=\mathcal{H}_{sc}+\mathcal{H}_{em}+\mathcal{H}_{int}
\end{align}
In the long-wavelength regime relevant to this work, where the wavelength of the electromagnetic field is much larger than the superconducting coherence length, the condensate can be effectively modelled as a charged bosonic gas \cite{SchriefferTheorySuperconductivity1964} in a finite volume ($V$), described by the Hamiltonian density of a non-relativistic complex scalar field \cite{L.BQuantumFieldTheory2014}:
\begin{align}
\begin{split}
    \mathcal{H}_{sc}&=\frac{\hbar^2}{m}\nabla\psi^\dagger(x,t)\cdot \nabla \psi(x,t)\\&+V\kappa\left(\psi^\dagger(x,t)\psi(x,t)\right)^2-2\mu\psi^\dagger(x,t)\psi(x,t)
\end{split}
\end{align}
The first term represents the kinetic energy of the condensate in terms of its field operator $\psi(x,t)$, while the remaining terms represent its potential energy. The second term describes the self-interaction of the condensate, governed by the interaction constant $\kappa$. Given that single-particle excitations of this field are defined here as Cooper pairs, this interaction constant is defined analogously to that of the BCS theory \cite{PhysRev.108.1175}. The final term normalizes the energy of the system to its chemical potential, $\mu$ \cite{TinkhamIntroductionSuperconductivity2015}. For consistency, we adopt SI units throughout the derivation.

The electromagnetic field is described by the following Hamiltonian density, under the Coulomb gauge:
\begin{gather}
    \mathcal{H}_{em}=\frac{\epsilon_o}{2}\left(\partial_o\mathbf{A}(x,t)\right)^2+\frac{1}{2\mu_o}\left(\nabla\times \mathbf{A}(x,t)\right)^2
\end{gather}

The interaction Hamiltonian density is derived via minimal coupling, using the condensate's induced current density obtained from Noether's theorem by the conservation of its charge:
\begin{align}
\begin{split}
    \mathcal{H}_{int}&=-i\mathbf{A}(x,t)\frac{q\hbar}{m}\Big[\nabla\psi(x,t) \psi^\dagger(x,t)\\
    &\qquad\qquad\qquad\qquad\qquad-\nabla\psi^\dagger(x,t)\psi(x,t)\Big]
\end{split}
\end{align}
The first term represents the current density of the negatively charged condensate, while the second term represents the current density associated with regions that are positively charged due to its absence \cite{L.BQuantumFieldTheory2014}. Higher-order terms proportional to $|A|^2$, which appear in the full gauge-invariant formulation, are neglected here under the assumption that the applied vector potential is sufficiently weak, as discussed further in the Supplement \cite{supp}.

We quantize the condensate and electromagnetic field separately, by expressing their constituent fields as
\begin{gather}
    \psi\left(x,t\right)=\frac{1}{\sqrt{V}}\sum_{k}{\hat{u}_k e^{i\left(\omega_k t+\mathbf{k} \cdot \mathbf{x}\right)}}\label{eqn:scFieldOperator1}
\end{gather}
\begin{gather}
    \mathbf{A}\left(x,t\right)=\sum_{\lambda}{\sqrt{\frac{\hbar}{2V\epsilon_o \omega_{\lambda}}}\boldsymbol{\epsilon}_{\lambda}}\hat{a}_{\lambda}e^{i\left(\omega_{\lambda}t+\mathbf{k_\lambda}\cdot \mathbf{x}\right)}+\text{H.c.}
\end{gather}
Here, the superconducting condensate is described in terms of the operators $\hat{u}_k$ and their Hermitian conjugates $\hat{u}_k^\dagger$, which represent the annihilation and creation of Cooper pairs with wavevector $k$, respectively. Although these operators retain the underlying fermionic structure and do not strictly obey bosonic commutation relations, in the long-wavelength limit ($q\approx0$) they reproduce the collective condensate mode and behave effectively as bosons. We consider electromagnetic field modes with frequencies below the superconducting gap ($\omega_{\lambda}<2\Delta/\hbar$), labelled by the index $\lambda$. Using the expansions above, the resulting quantized Hamiltonian is given in Equation (\ref{Eqn:quantizedHamiltonian}), with the full derivation provided in the Supplement \cite{supp}.

\begin{widetext}
    \begin{multline}
    \label{Eqn:quantizedHamiltonian}
    H=\sum_{k,\lambda,q}{} 2\left(\hbar \omega_k-\mu\right)\hat{u}_k^\dagger\hat{u}_k+\kappa'\hat{u}_k^\dagger\hat{u}_{k\pm q} +\frac{\hbar\omega_{\lambda}}{2}\left(\hat{a}_{\lambda}\hat{a}_{\lambda}^\dagger + \hat{a}_{\lambda}^\dagger\hat{a}_{\lambda}\right)
    +\sqrt{\frac{q^2 c^2\hbar}{V\epsilon_o\omega_{\lambda}}}\left(\hat{a}_{\lambda}e^{i\omega_{\lambda}t}+\hat{a}^\dagger_{\lambda}e^{-i\omega_{\lambda}t}\right)\left(\hat{u}_k \hat{u}_k^\dagger+\hat{u}_k^\dagger \hat{u}_k\right)
    \end{multline}
\end{widetext}

The first two terms of the quantized Hamiltonian correspond to the BCS Hamiltonian \cite{PhysRev.108.1175}, expressed in terms of $\hat{u}_k$ and $\hat{u}_k^\dagger$. The quartic interaction term in the underlying field theory has been simplified via the Bogoliubov mean-field approximation (see Supplement \cite{supp}), reducing it to an effective quadratic form \cite{Bogolyubov1947}. In this approximation, the macroscopic condensate amplitude ($n_0$) is absorbed into the effective interaction coefficient ($\kappa' = \kappa n_0$), which is determined self-consistently via the BCS gap equation. The third term describes the free Hamiltonian of the quantized EM field modes \cite{G.CQuantumOptics2014}. The final term is the quantized representation of their interaction, shown here in momentum-space after applying the dipole approximation, and describes the coupling energy between a Cooper pair with wavevector $k$ and a photon mode with wavevector $k_\lambda$.

\begin{figure*}
    \centering
    \includegraphics[width=0.80\textwidth]{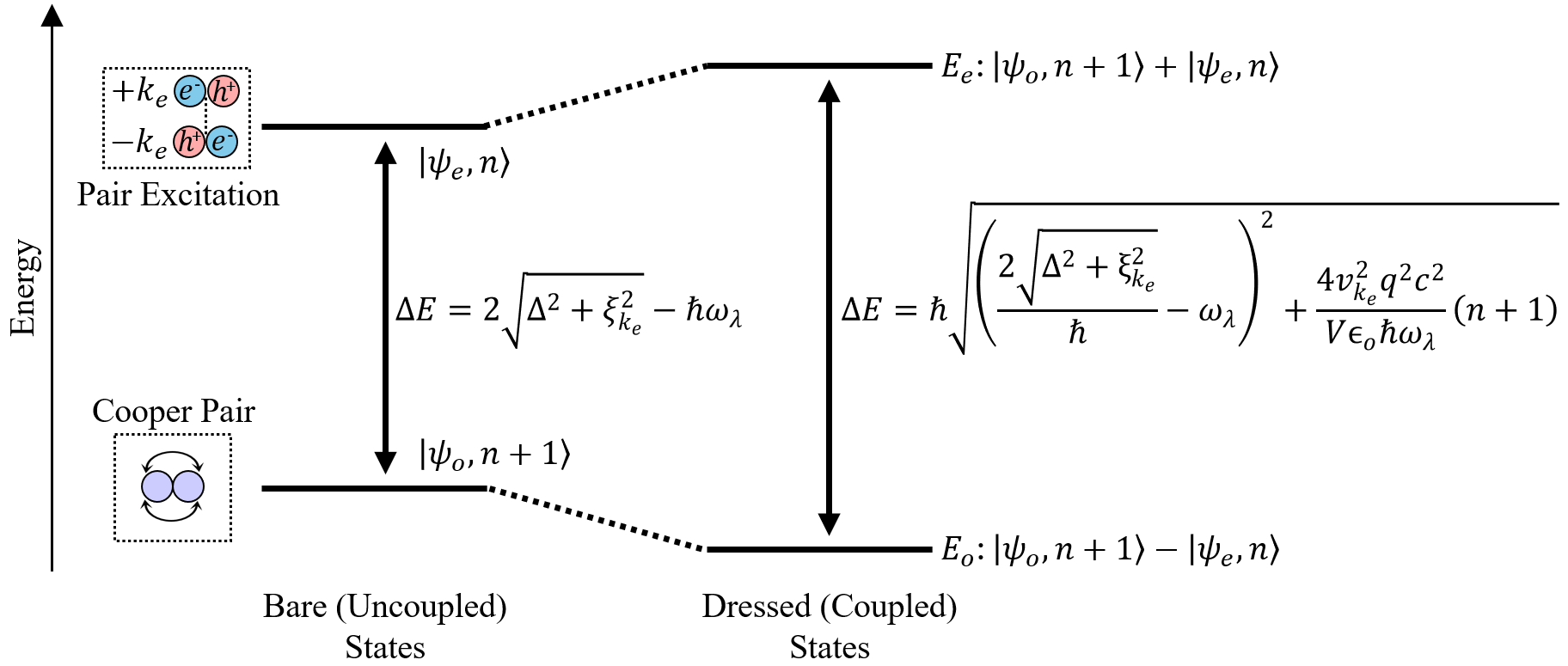}
    \caption{Schematic illustration of photon-Cooper pair coupling in a superconducting condensate. Left: Bare (uncoupled) states consisting of the BCS ground state ($\psi_o$) with $n+1$ photons, and a pair-excited state ($\psi_e$) with $n$ photons. The excitation corresponds to the promotion of a Cooper pair into its constituent electron and hole pairs at opposite momentum states, $+k_e$ and $-k_e$, resulting in a localized charge separation that interacts with the quantized electromagnetic field. Right: Dressed (coupled) states formed by coherent superpositions of the bare states, exhibiting an energy splitting enhanced by photon-Cooper pair entanglement, as described by Equation (\ref{eqn:vacRabiSplit}). Here, $\xi_{k_e}$ is the kinetic energy of the excited Cooper pair relative to the Fermi level, $\Delta$ is the superconducting gap, $v_{k_e}^2$ is the Cooper pair occupation probability at momentum $k_e$, and $V$ is the volume in which the condensate is confined.}
    \label{fig:energyDiagram}
\end{figure*}

To calculate the energy splitting associated with dressed-state formation, we first define the relevant excited state of the condensate. Since the condensate is modelled as a charged bosonic gas of Cooper pairs, we assume that the pairing is preserved during the photon interaction, allowing us to neglect spin-related effects \cite{L.BQuantumFieldTheory2014}. We therefore consider pair excitations \cite{PhysRev.108.1175}, in which a Cooper pair is promoted out of the condensate's collective wavefunction but remains bound through phonon coupling. In this regime, the condensate is treated using an effective bosonic description of these composite pair excitations, with their non-ideal commutation relation $([\hat{u}_k,\hat{u}_{k}^\dagger]=1-\hat{n}_{-k\downarrow}-\hat{n}^\dagger_{k\uparrow})$ explicitly taken into account. In the BCS framework, this corresponds to replacing the probabilistic Cooper pair amplitude (with occupation probability $v_k^2$) by a definite occupation of the $+k$ and $-k$ electron states and corresponding holes at $-k$ and $+k$, as illustrated in Figure \ref{fig:energyDiagram}. Although the system as a whole remains electrically neutral, this localized charge separation couples to the quantized electromagnetic field to form dressed states. These dressed states represent photon-Cooper pair entanglement in the bipartite system.

We model the excited state by starting from the BCS ground state and introducing a definite pair excitation at momentum $k_e$ using a creation operator, while all other $k$-states retain their BCS occupations:
\begin{align}
    \ket{\psi_e}=\prod_{k\neq k_e}\left(u_k+v_k\hat{u}_k^\dagger\right)\ket{0_k} \ket{1_{k_e}}
\end{align}
where, $\ket{1_{k_e}}=\hat{u}_{k_e}^\dagger\ket{0_{k_e}}$. The energy difference between this excited state and the ground state of the condensate is derived in the Supplement \cite{supp} and given by:
\begin{align}
        \hbar\omega_{21} & \stackrel{\triangle} = 4\Delta u_{k_e} v_{k_e} + 2\xi_{k_e} \left(1-2v_{k_e}^2\right)=2\sqrt{\Delta^2+\xi_{k_e}^2} 
\end{align}
The first term reflects the change in potential energy associated with the pair excitation, while the second term corresponds to the change in kinetic energy. This result agrees with the standard excitation energy predicted by BCS theory \cite{PhysRev.108.1175}.

This consistency is a consequence of defining the bare states in accordance with BCS theory. To incorporate field coupling, we now determine the eigenenergies of the full Hamiltonian (Equation (\ref{Eqn:quantizedHamiltonian})), and show that their difference still reduces to the BCS result in the appropriate limit. The corresponding eigenstates are dressed states, representing entangled superpositions of Cooper pair and photon excitations, whose energies are shifted due to this interaction. To compute these energies, we project the full Hamiltonian onto the two-level subspace spanned by the bare states, yielding an effective atomic Hamiltonian (see Supplement \cite{supp}).
\begin{align}
    H=\hbar\omega_{\lambda}n+\left(\frac{\hbar\omega_{21}-\hbar\omega_\lambda }{2}\right)\hat{\sigma} _z+\sqrt{\frac{v_{k_e}^2q^2c^2\hbar}{V\epsilon_o\omega_{\lambda}}}\sqrt{n+1}\hat{\sigma}_x
\end{align}
Here, the coupling between the condensate and the electromagnetic field introduces off-diagonal terms into the atomic Hamiltonian, allowing us to redefine the system's basis and obtain dressed states whose energies correspond to its eigenvalues. The difference between these eigenvalues defines the renormalized excitation energy of the coupled system:
\begin{align}
    \Delta E=\hbar\sqrt{\left(\frac{2\sqrt{\Delta^2+\xi_{k_e}^2}}{\hbar}-\omega_{\lambda}\right)^2+\frac{4v_{k_e}^2q^2c^2}{V\epsilon_o\hbar\omega_{\lambda}}\left(n+1\right)} \label{eqn:vacRabiSplit}
\end{align}
Here, $\Omega=\Delta E/\hbar$ is the renormalized excitation frequency of the system, which depends explicitly on the photon occupation number, $n$. This result is analogous to $n$-photon Rabi oscillations for $n>0$, and to vacuum Rabi oscillations when $n=0$, as in cavity QED, with $\Omega$ representing the corresponding Rabi frequency in either case.

In the large-volume limit, the energy corresponding to transitions between these dressed states for Cooper pairs at the Fermi surface ($k_e=k_F$) reduces to,
\begin{align}
    \Delta E|_{k_e=k_F,V\rightarrow \infty}=2\Delta-\hbar\omega_{\lambda}
\end{align}
In this limit, the coupling term vanishes, and the excitation energy approaches the minimum predicted by BCS theory, confirming the consistency of our result with the conventional representation.

\begin{figure}
    \centering
    \includegraphics[width=0.48\textwidth]{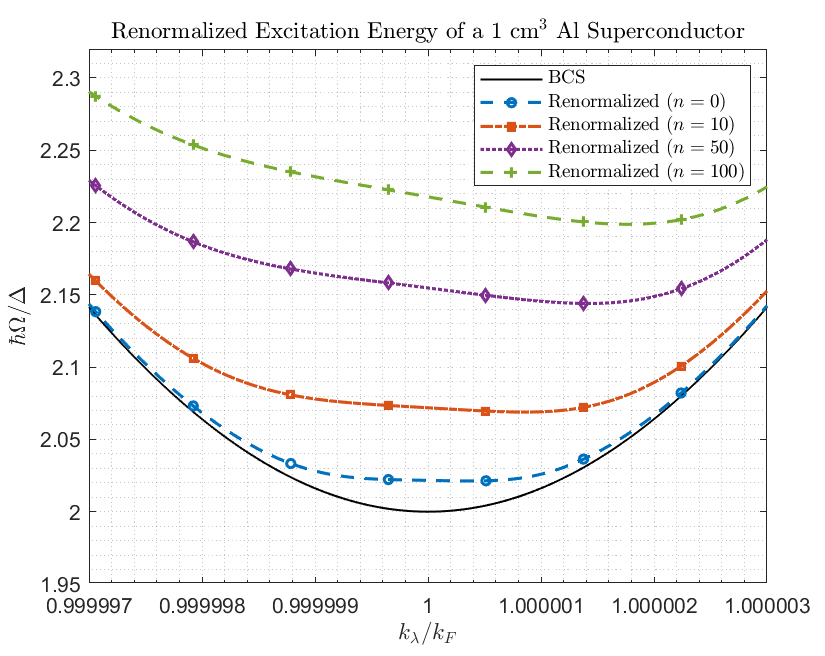}
    \caption{Single-photon Cooper pair excitation energy ($\hbar \Omega$) for a 1 cm$^3$ aluminum superconductor, calculated from Equation (\ref{eqn:vacRabiSplit}). The energy is shown as a function of photon number $n$, for a field frequency chosen near the pair-breaking energy of the superconducting gap ($\omega_\lambda=2\Delta/\hbar$). The result is plotted near the Fermi level ($k_F$) and normalized to the gap energy ($\Delta$), with the large-volume limit ($V\rightarrow\infty$) used to define the BCS baseline.}
    \label{fig:Renormalized_Excitation_Energy.png}
\end{figure}

% Results
A key implication of this result is that the energy required to excite a Cooper pair in the presence of an electromagnetic field now depends on three main factors: the number of coupled photons, their frequency, and the sample's volume. To illustrate the magnitude of this effect, consider a type I superconductor such as aluminum \cite{RobertsBenjaminWashingtonPropertiesSelectedSuperconductive1978}, occupying a 1 cm$^3$ volume. Assuming the superconducting condensate uniformly fills the sample volume \cite{TinkhamIntroductionSuperconductivity2015}, we estimate an upper limit for the number of confined photons using the EM mode density: $n=\left(\frac{8\pi}{3}\right)\left(\frac{E}{hc}\right)^3V\approx 100$. We evaluate the photon frequency near the pair-breaking energy of Cooper pairs at the Fermi surface, $\omega_\lambda\approx 2\Delta/\hbar$, as this energy gives the largest deviation from BCS theory and therefore provides a first-order estimate of the effect's magnitude. Although the true vacuum response spans a broad, continuous range of frequencies, focusing on this dominant contribution offers a conservative order-of-magnitude estimate. As shown in Figure \ref{fig:Renormalized_Excitation_Energy.png}, under these conditions, the superconducting gap ($2\Delta$) increases by approximately $0.22\Delta$, corresponding to an energy shift of 9.6 GHz relative to the prediction of BCS theory \cite{PhysRev.108.1175}.

While we derive this result for an isolated superconductor, its practical implications can be seen by considering its effect on the critical current of a Josephson junction. Using the Ambegaokar-Baratoff relation, $I_c=\frac{\pi\Delta(0)}{2eR_n}$ \cite{TinkhamIntroductionSuperconductivity2015}, we can estimate the enhancement for an aluminum junction with a normal-state resistance of 100 $\Omega$. In this scenario, the critical current increases from $2.8$ $\mu$A to $3.1$ $\mu$A, representing an increase of $300$ nA ($11\%$).

Thus, our theory establishes a mechanism for microwave-enhanced superconductivity in the quantum regime, arising from the formation of dressed states of Cooper pairs. This is in contrast to the conventional Eliashberg framework \cite{osti_4687835, Eliashberg1971}, which describes microwave enhancement as a non-equilibrium effect arising from the redistribution of quasiparticle excitations under high-intensity classical electromagnetic fields \cite{C.SNonequilibriumSuperconductivity1978}. In that approach, the field perturbs the quasiparticle distribution near the energy gap, promoting Cooper pair formation beyond the critical conditions \cite{C.SNonequilibriumSuperconductivity1978}. In contrast, our theoretical framework is valid for arbitrary photon numbers and applies universally to any mesoscopic superconducting system.

% vacuum fluctuations
So far, we have focused on how the quantized electromagnetic field modifies the properties of the superconducting condensate due to their interaction. However, due to the entanglement between the two, the condensate must also influence the quantum optical properties of the confined electromagnetic field. To understand the significance of this, we first note that the superconducting gap is modified even without an external photon source ($n=0$), as shown in Equation (\ref{eqn:vacRabiSplit}). This modification arises from the coupling of Cooper pairs to virtual photons in the vacuum fluctuations of the electromagnetic field, resulting in a non-negligible energy shift \cite{G.CQuantumOptics2014}. As a first-order approximation, we evaluate the interaction at a representative frequency near the pair-breaking energy ($\omega_\lambda=\omega_{21}$), while recognizing that a broad continuum of virtual-photon frequencies in the vacuum field contributes to the full effect. For the aluminum sample previously discussed, we calculate a renormalized minimum excitation energy of $2.02\Delta$ (0.87 GHz), as shown in Figure \ref{fig:Renormalized_Excitation_Energy.png}. Although this shift lies within typical experimental uncertainty for BCS gap measurements, it is notable given that the sample is macroscopic (1 cm$^3$) and that the response increases as its volume decreases.

We now explicitly evaluate how the presence of the condensate modifies field fluctuations by computing their variance. Within the minimal coupling approximation, only the electric field variance is modified by the condensate. This is because, unlike the magnetic field, the total electric field includes contributions both from the incident wave and from the induced current density:
\begin{align}
    \mathbf{E}_T=\mathbf{E}_{inc}+\mathbf{E}_{J}
\end{align}
We derive the incident component from the earlier definition of the quantized magnetic vector potential:
\begin{align}
    \begin{split}
    \mathbf{E}_{inc}=-\sum_{\lambda}{i\sqrt{\frac{\hbar\omega_{\lambda}}{2V\epsilon_o}}\boldsymbol{\epsilon}_{\lambda}}\hat{a}_{\lambda}e^{i\left(\omega_{\lambda}t+\mathbf{k_\lambda}\cdot \mathbf{x}\right)} + \text{H.c.}\label{eqn:electricFieldQuantized}
    \end{split}
\end{align}
The induced component is obtained from Ampere's law, expressed in terms of the condensate's current density:
\begin{align}
    \begin{split}
    \mathbf{E}_J&=\sum_{k,\lambda}{-\frac{icq}{V\omega_{\lambda}\epsilon_o}}\left(\hat{u}_k^\dagger\hat{u}_k-\hat{u}_k\hat{u}_k^\dagger\right)\\
    &\qquad\qquad+\frac{ikq\hbar}{V\omega_{\lambda}\epsilon_om}\left(\hat{u}_k\hat{u}_k^\dagger+\hat{u}_k^\dagger\hat{u}_k\right)
    \end{split}
\end{align}
As shown in the Supplement \cite{supp}, the variance of the total electric field in the presence of the ground state of the condensate is then given as,
\begin{align}
    \left(\Delta\mathbf{E}\right)^2&=\braket{\psi_o,n|\mathbf{E}_T^2|n,\psi_o}-\braket{\psi_o,n|\mathbf{E}_T|n,\psi_o}^2\nonumber\\
    &=\sum_{k,\lambda}{}\frac{\hbar\omega_{\lambda}}{V\epsilon_o}\left(n+\frac{1}{2}\right)-\left(\frac{cq}{V\omega_{\lambda}\epsilon_o}\right)^2\left(\frac{\Delta^2}{\Delta^2+\xi_k^2}\right) \label{eqn:fieldvariance}
\end{align}

Here, we find that electric field fluctuations are suppressed in the presence of the condensate, due to the localized charge separation between electrons and holes following the excitation of Cooper pairs. This behaviour is analogous to the suppression of fluctuations in dielectric media caused by induced polarization, as described by Glauber and Lewenstein \cite{G.LQuantumOpticsDielectric1991}. In fact, in the limiting case where the condensate is either uncharged ($q=0$) or non-superconducting ($\Delta=0$), our variance expression reduces to the free-space result derived in their work \cite{G.LQuantumOpticsDielectric1991}, as expected, given that this either corresponds to the absence of the minimal coupling interaction or the condensate itself.

The magnitude of this fluctuation suppression is governed by the factor $\Delta^2/\left(\Delta^2+\xi_k^2\right)$, which effectively quantifies the coherence of the superconducting state. As the temperature increases, this ratio decreases both due to the increasing kinetic energy ($\xi_k$) of the Cooper pairs as well as the reduction in the superconducting gap ($\Delta$). These effects lead to delocalized thermal excitations, which do not oscillate coherently with the applied field and therefore do not contribute to the formation of dressed states.

To directly quantify how the gap is modified by an electric field with this fluctuation magnitude, we consider a semiclassical approximation in which the quantized Cooper pair system interacts with a classical EM field. In this model, the interaction Hamiltonian is given by
\begin{align}
    H_{int}=\mathbf{A}qc\sqrt{2}(\hat{u}_k\hat{u}_k^\dagger+\hat{u}_k^\dagger\hat{u}_k)
\end{align}
The corresponding renormalized excitation energy becomes:
\begin{align}
    \Delta E_{\text{S-Cl}}=\hbar\sqrt{ \left(\frac{2\sqrt{\Delta^2+\xi_{k_e}^2}}{\hbar}-\omega_{\lambda} \right)^2+\frac{8v_{k_e}^2q^2c^2}{\hbar^2\omega_{\lambda}^2}|\mathbf{E}|^2}. \label{eqn:ATResonance}
\end{align}
For the aluminum superconductor described previously, we estimate the vacuum electric field fluctuations near the pair-breaking frequency to have a standard deviation of $1.8$ mV/m, using Equation (\ref{eqn:fieldvariance}), with a corresponding minimum excitation energy of $2.015\Delta$, as per Equation (\ref{eqn:ATResonance}). This result agrees with the fully quantized approach, to within $0.3\%$, and illustrates how the superconducting condensate alters the quantum properties of the electromagnetic vacuum.

In conclusion, we have developed a quantum field theoretical framework for electromagnetic field quantization in the presence of a superconducting condensate, modelled as a charged bosonic gas. By reformulating the system's Hamiltonian via minimal coupling into an effective two-level model, we show that photon-Cooper pair entanglement gives rise to microwave-dressed states near the Fermi surface. This interaction renormalizes the superconducting gap and enables microwave-enhanced superconductivity in the quantum regime. Additionally, we demonstrate that the condensate exerts a back-action on the electromagnetic field, suppressing electric field fluctuations, including those originating from the vacuum, thus revealing new aspects of the equilibrium quantum electrodynamics of superconductors. Given the shared field-theoretical foundation of our model and Landau-Ginzburg theory, these results are expected to extend naturally to Type II superconductors by incorporating vortex states and spatial variations.

The authors acknowledge the support of the Natural Sciences and Engineering Research Council of Canada (NSERC), the Discovery Grant program, RGPIN-2024-04522.  

\nocite{V.T.TPrinciplesSuperconductiveDevices1981}
\bibliography{references.bib}

\end{document}